\documentclass[letters,usenatbib]{mnras}

\usepackage{newtxtext,newtxmath}
\usepackage[T1]{fontenc}

\DeclareRobustCommand{\VAN}[3]{#2}
\let\VANthebibliography\thebibliography
\def\thebibliography{\DeclareRobustCommand{\VAN}[3]{##3}\VANthebibliography}

\usepackage{graphicx}	% Including figure files
\usepackage{amsmath}	% Advanced maths commands
\title[Resolving the Spiderweb protocluster]{Spider-Webb: JWST Near Infrared Camera resolved galaxy star formation and nuclear activities in the Spiderweb protocluster at z = 2.16}

\author[R. Shimakawa et al.]{Rhythm Shimakawa,$^{1,2}$\thanks{E-mail: rhythm.shimakawa@aoni.waseda.jp (RS)}
Yusei Koyama,$^{3,4,5}$
Tadayuki Kodama,$^{6}$
Helmut Dannerbauer$^{7,8}$
\newauthor
J. M. P\'{e}rez-Mart\'{i}nez,$^{7,8}$
Huub J. A. R\"ottgering,$^{9}$
Ichi Tanaka,$^{4}$
Chiara D'Eugenio,$^{7,8}$
\newauthor
Abdurrahman Naufal,$^{5}$
Kazuki Daikuhara$^{6}$
and Yuheng Zhang$^{7,8,10,11}$
\\
$^{1}$Waseda Institute for Advanced Study (WIAS), Waseda University, 1-21-1, Nishi-Waseda, Shinjuku, Tokyo 169-0051, Japan\\
$^{2}$Center for Data Science, Waseda University, 1-6-1, Nishi-Waseda, Shinjuku, Tokyo, 169-0051, Japan\\
$^{3}$National Astronomical Observatory of Japan (NAOJ), National Institutes of Natural Sciences, 2-21-1 Osawa, Mitaka, Tokyo 181-8588, Japan\\
$^{4}$Subaru Telescope, National Astronomical Observatory of Japan, National Institutes of Natural Sciences, 650 North A'ohoku Place, Hilo, HI 96720, USA\\
$^{5}$Department of Astronomical Science, The Graduate University for Advanced Studies, 2-21-1 Osawa, Mitaka, Tokyo 181-8588, Japan\\
$^{6}$Astronomical Institute, Tohoku University, 6-3, Aramaki, Aoba, Sendai, Miyagi 980-8578, Japan\\
$^{7}$Instituto de Astrof\'{\i}sica de Canarias, E-38205 La Laguna, Tenerife, Spain\\
$^{8}$Universidad de La Laguna, Dpto. Astrof\'{i}sica, E-38206 La Laguna, Tenerife, Spain\\
$^{9}$Leiden Observatory, Leiden University, PO Box 9513, NL-2300 RA Leiden, the Netherlands\\
$^{10}$Purple Mountain Observatory, Chinese Academy of Sciences, 10 Yuanhua Road, Nanjing, 210023, China\\
$^{11}$School of Astronomy and Space Science, University of Science and Technology of China, Hefei, Anhui 230026, China\\
}

\date{Accepted 2024 October 14. Received 2024 October 13; in original form 2024 September 8}
\pubyear{2024}

\begin{document}
\label{firstpage}
\pagerange{\pageref{firstpage}--\pageref{lastpage}}
\maketitle

\begin{abstract}
Near-infrared (NIR) emission is less affected by dust than UV and optical emission and is therefore useful for studying the properties of dust-obscured galaxies. 
Although rest-frame NIR observations of high-redshift galaxies have long been made using space telescopes, their structures were unresolved due to the lack of angular resolution.
This letter reports the early results from the analysis of high-resolution Pa$\beta$ imaging of the Spiderweb protocluster at $z=2.16$ with the JWST Near Infrared Camera.
We investigate radial profiles of Pa$\beta$ lines and rest-frame NIR continua from luminous H$\alpha$-emitting galaxies (HAEs) in the protocluster. 
Particularly, we compare those of 11 HAEs (N-HAEs) on the star-forming main sequence with those of 8 HAEs (X-HAEs) with X-ray AGNs. 
Resultant composite Pa$\beta$ line images of N-HAEs indicate significant star formation in galactic disks. 
In contrast, X-HAEs are dominated by point source components rather than outer star formation, as inferred from our earlier work based on multi-wavelength SED fitting.
Given their higher stellar potentials suggested from their rest-frame NIR images, the different characteristics may be driven by the impact of AGN feedback.
\end{abstract}

\begin{keywords}
galaxies: clusters: individual: PKS~1138-262 -- galaxies: evolution -- galaxies: formation -- galaxies: high-redshift
\end{keywords}

%%%%%%%%%%%%%%%%%%%%%%%%%%%%%%%%%%%%%%%%%%%%%%%%%%

%%%%%%%%%%%%%%%%% BODY OF PAPER %%%%%%%%%%%%%%%%%%

\section{Introduction}
\label{s1}

The redshift interval of $z=$ 2--4 is known as the peak epoch of cosmic star formation rate (SFR) density, providing a key to understanding galaxy formation and evolution (e.g., \citealt{Madau1996,Lilly1996,Hopkins2006,Madau2014}, and references therein).
While significant progress has been made over the past decades, a framework of regulation mechanisms of star formation in massive galaxies remains controversial (e.g., \citealt{Shapley2011,Silk2012,Schawinski2014,Somerville2015,Naab2017}).
Active galactic nucleus (AGN) feedback (e.g., \citealt{Silk1998,Sijacki2007,Fabian2012,ForsterSchreiber2020}) is currently the most successful solution to explain observations, such as stellar mass to halo mass relation in the bulk of modern hydrodynamic simulations (e.g., \citealt{Dubois2014,Dave2019,Terrazas2020,Wellons2023}). 

Thus, understanding the intricate interplay (co-evolution) between galaxies and supermassive black holes (SMBHs) has become a central theme in the latest research on galaxy formation and evolution. 
Modern hydrodynamic simulations commonly implement at least two channels (so-called radiative and kinetic modes) in AGN feedback (e.g., \citealt{Schaye2015,Weinberger2017}), and some model suites also include other mechanisms, such as X-ray heating \citep{Choi2012,Dave2019} and cosmic ray feedback \citep{Wellons2023,Hopkins2023}.
These simulations suggest that AGN feedback at low accretion rates in SMBHs is considered important for massive galaxies, as it dominates the time-integrated feedback energy.
In addition, unravelling AGN feedback mechanisms is tightly linked to understanding BH accretion mechanisms following a well-known M$_\mathrm{BH}$--$\sigma$ relation \citep{Magorrian1998,Ferrarese2000,King2003,Kormendy2013,Heckman2014}.
Otherwise, over-(under-)massive black holes result in over-(under-)quenching in the stellar mass to halo mass relation \citep{Wellons2023}, requiring further adjustments to feedback models. 

However, investigating solely luminous AGNs may not contribute significantly to understanding AGN feedback because it only traces a short period of AGN activities, and because energy contributions from radiative AGN feedback at high accretion states may be fairly low compared with the total budget of energy injections \citep{Terrazas2020,Piotrowska2022,Ward2022,Bluck2023,Bluck2024,Hartley2023}. 
Weak signs of AGNs have been frequently reported in quiescent and post star-forming galaxies through deep observations or stacking analyses (e.g., \citealt{Kriek2009,Olsen2013,Marsan2015,Man2016,Gobat2017,Ito2022,Kubo2022,Carnall2023,Belli2023,Shimakawa2024,Lopez2024,Park2024,Bugiani2024}).
However, a systematic study of post-star-forming galaxies that host SMBHs at moderate-to-low accretion rates is quite expensive, since deep and wide-field observations with various facilities in multi-wavelength are required. 

An efficient strategy for addressing such an observational challenge is to target maturing protoclusters at $z=$ 2--4, where a number of massive galaxies under star formation quenching can be observed at a lower cost owing to a smaller number of pointings. 
One of the best-studied protoclusters, the Spiderweb protocluster \citep{Carilli1997,Pentericci1997,Pentericci2000,Kurk2000}, would be a good testbed for examining AGN feedback on massive galaxies.
Based on the X-ray to the sub-mm data, \citet[S24 hereafter]{Shimakawa2024} reported that a third of massive H$\alpha$ emitters (HAEs) are undergoing a transition from star-forming to quiescent, and significant fractions of their H$\alpha$+[N{\sc ii}] lines would originate from low-luminosity AGNs like low ionisation nuclear emission-line regions (LINERs; \citealt{Heckman1980}) rather than star formation. 
However, the results were inconclusive as they heavily relied on complex spectral energy distribution (SED) modelling; therefore, more direct verification is required to resolve star formation and AGN activities in these massive HAEs. 

Given such a backdrop, this letter examines radial profiles of Pa$\beta$ lines of 19 massive HAEs with M$_\star>2\times10^{10}$ M$_\odot$ in the Spiderweb protocluster at $z=2.16$, based on newly obtained JWST/NIRCam F405N \& F410M images \citep{Shimakawa2024b,Perez-Martinez2024}.
Pa$\beta$ gives us with a less dust-biased view compared to H$\alpha$ \citep{Kennicutt1998,Calzetti2007} and its wavelength redshifted to the protocluster is just covered by the F405N filter.
We select the targets from S24 (table~A), which have measured stellar masses and star formation rates (SFRs) for 84 HAEs based on the multi-wavelength photometry from X-ray to submm data, including 14 HAEs with X-ray counterparts \citep{Tozzi2022a}.
This work is motivated to resolve star formation and AGN activities for these massive HAEs at $\sim$ one kpc resolution with NIRCam, enabling us to validate the previous arguments. 
Throughout the letter, we assume a flat lambda cold dark matter model with $h=0.693$ and $\Omega_M=0.286$, which are consistent with those obtained from the WMAP nine-year data \citep{Hinshaw2013}.
We use the \citet{Chabrier2003} stellar initial mass function (IMF) and the AB magnitude system \citep{Oke1983}.
When we refer to figures or tables in this letter, we designate their initials by capital letters (e.g., Fig.~1 or Table~1) to avoid confusion with those in the literature (e.g., fig.~1 or table~1).

\section{Targets and data}
\label{s2}

% figure 1
\begin{figure}
\centering
\includegraphics[width=0.95\columnwidth]{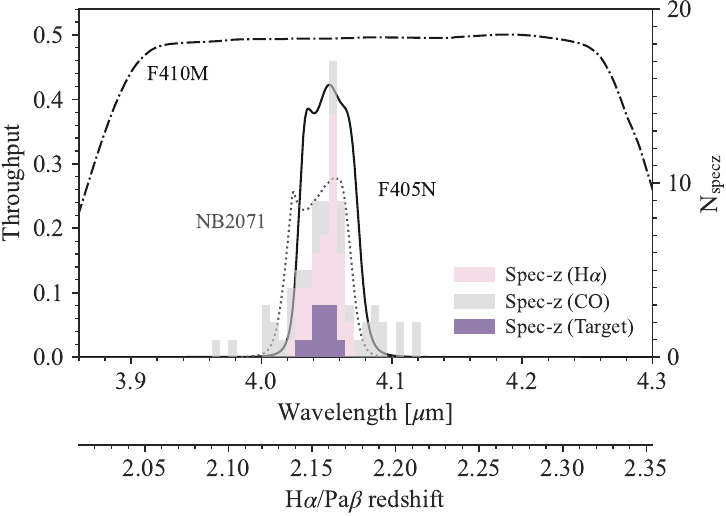}
\caption{
Filter response curves and redshift distribution of spec-$z$ members in the Spiderweb protocluster.
The solid, dash-dot, and dotted lines indicate the filter throughput of the NIRCam/F405N, F410M, and MOIRCS/NB2071, respectively.
Spec-$z$ confirmations with H$\alpha$ and CO($1-0$) lines are shown separately by pink and grey \citep{Shimakawa2014,Perez-Martinez2023,Jin2021}.
Moreover, if available, spec-$z$ distribution of our targets (N-HAEs and X-HAEs) is shown by the magenta histogram.
}
\label{fig1}
\end{figure}

Our primary goal is to study radial profiles of Pa$\beta$ lines of massive HAEs with X-ray counterparts in S24, where X-ray sources are observed using a deep survey with the Chandra X-ray observatory \citep{Tozzi2022a}. 
Fig.~\ref{fig1} shows the redshift distributions of our targets, if their spectroscopic redshifts are available, and known protocluster members with spectroscopic confirmations by previous studies on H$\alpha$ and CO lines \citep{Shimakawa2014,Jin2021,Perez-Martinez2023}. 
It indicates the target HAEs are adequately covered by the both narrowband filters, NB2071 on Subaru/MOIRCS in H$\alpha$ line \citep{Shimakawa2018b} and F405N on the JWST/NIRCam \citep{Rieke2005} in Pa$\beta$ line, enabling us to analyse their Pa$\beta$ line properties with the F405N image.

% figure 2
\begin{figure}
\centering
\includegraphics[width=0.9\columnwidth]{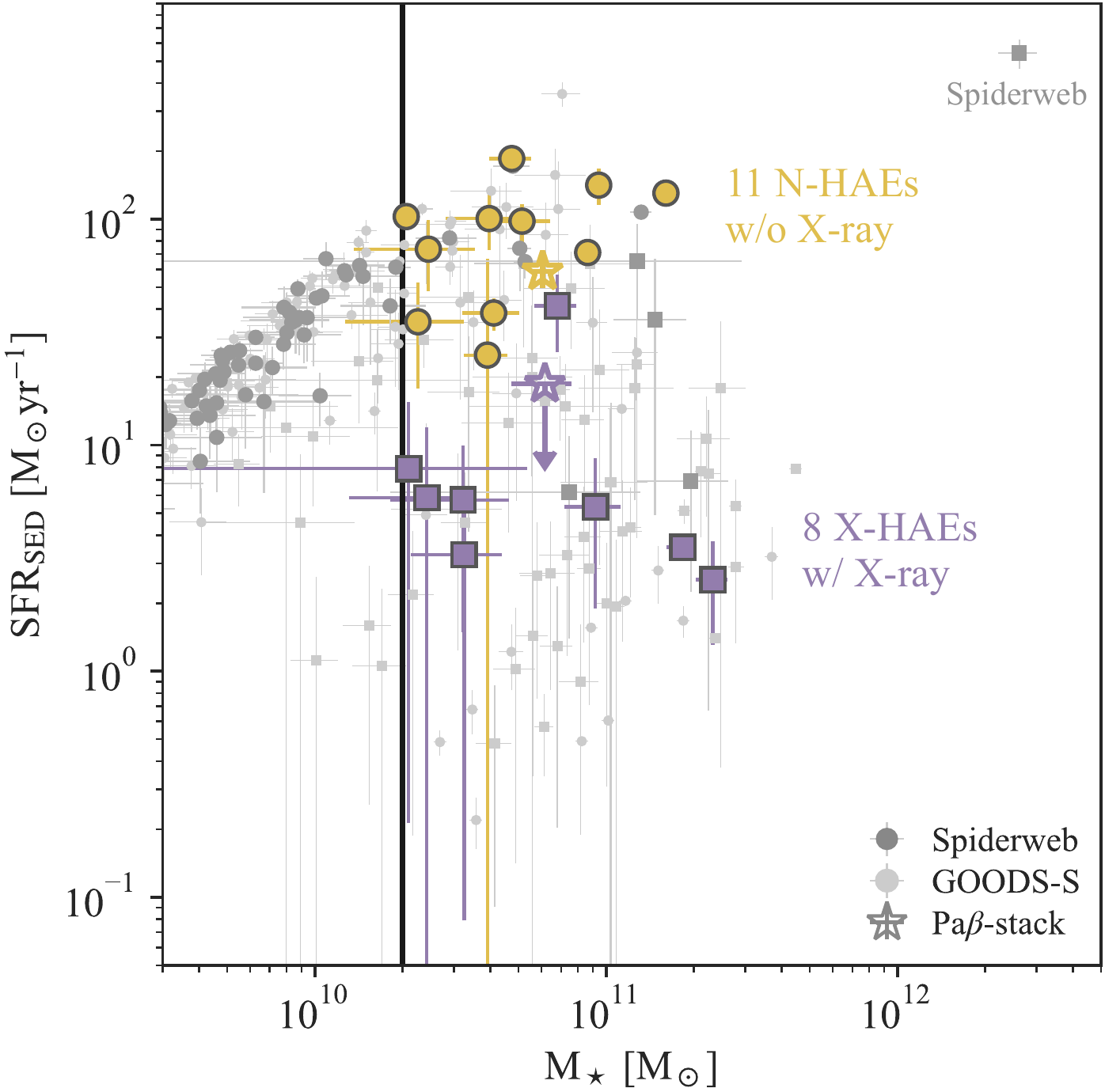}
\caption{
SED-based SFRs versus stellar masses for HAE members of the Spiderweb protocluster (coloured symbols) and reference samples from the GOODS-S field at $z=$ 1.5--2.5 (grey symbols) in S24.
Galaxies with and without X-ray counterparts are represented by circles and squares, respectively. 
We adopt 11 N-HAEs (yellow) and 8 X-HAEs (magenta) with M$_\star>2\times10^{10}$ M$_\odot$ covered by the NIRCam imaging.
The star symbols are based on the stacking analysis, where their stellar masses are reproduced from median flux densities and their SFRs are based on dust-corrected Pa$\beta$ fluxes (see Section~\ref{s3}).
}
\label{fig2}
\end{figure}

Fig.~\ref{fig2} shows SED-inferred SFRs (SFR$_\mathrm{SED}$) averaged over the last 100 Myr and stellar masses of HAEs in S24 (fig.~4), which are derived from the {\sc x-cigale} SED fitting code \citep{Boquien2019,Yang2020,Yang2022} based on the multi-band photometry from various pieces of literature \citep{Kurk2000,Miley2006,Kodama2007,Seymour2007,Koyama2013a,Valtchanov2013,Dannerbauer2014,Shimakawa2018b,Tozzi2022a}.
S24 have suggested that a large fraction of X-ray detected HAEs consists of post star-forming galaxies well below the star-forming main sequence \citep{Brinchmann2004,Noeske2007,Daddi2007,Elbaz2007,Salim2007} despite them having H$\alpha$ (+[N{\sc ii}]) line emissions similar to those of HAEs without X-ray counterparts at a similar stellar mass range.
They also report that most of them host low-luminosity X-ray AGNs with $L_\mathrm{X,2-10keV}\lesssim4\times10^{43}$ erg~s$^{-1}$ that cannot be explained by star formation \citep{Tozzi2022a}.
Based on these previous studies, this letter focuses on 19 massive HAEs with stellar masses of $>2\times10^{10}$ solar mass (M$_\odot$), except for the Spiderweb radio galaxy, among 58 HAEs covered by our JWST/NIRCam images (Shimakawa et al. 2024b, fig.~2). 
They are divided into 8 HAEs with X-ray counterparts and 11 HAEs without X-ray detection for comparison (Fig.~\ref{fig2}).
The identification numbers in the respective samples are $[40,46,48,55,58,71,77,83]$ with X-ray detection and $[9,15,22,28,30,35,39,54,80,82,93]$ without X-ray detection (S24, table~A).\footnote{ID=28 was regarded as an X-HAE in \citet{Tozzi2022a}; S24, whereas Naufal et al. (submitted) noted that its neighbouring source is the actual X-ray counterpart. Another N-HAE (ID=89) is also within the FoV, but was excluded from the sample as it is likely to be a background emitter (S24).}
For convenience, these two samples are referred to as `X-HAEs' and `N-HAEs' hereafter.
Each of them exhibits similar properties in SFR$_\mathrm{SED}$ and stellar mass within each sample, except an X-HAE (ID=58) showing significantly higher SFR than the other X-HAEs (Fig.~\ref{fig2}).
However, it should be noted that the conclusions of this paper remain the same even if this X-HAE is not considered.
In that case, the central Pa$\beta$ component of the composite Pa$\beta$ radial profile in X-HAEs (Section~\ref{s3}) becomes slightly weaker.

The Pa$\beta$ line image (F405N) and corresponding continuum image (F410M) are obtained by the JWST/NIRCam at the long wavelength channel through the Cycle-1 GO~1572 program, in parallel with F115W and F182M imaging at the short wavelength channel \citep{Shimakawa2024b,Perez-Martinez2024}. 
The total integrations are 63 min and 21 min, respectively.
The typical $3\sigma$ flux limit in the narrowband selection is $\sim2\times10^{-18}$ erg~s$^{-1}$cm$^{-2}$, which is comparable with the H$\alpha$+[N{\sc ii}] flux limit of $\sim3\times10^{-17}$ erg~s$^{-1}$cm$^{-2}$ in the Subaru/MOIRCS H$\alpha$ imaging \citep{Shimakawa2018b}, under the dust-free line ratio of H$\alpha$/Pa$\beta=17.5$ in the case B recombination \citep{Luridiana2015}.
Details of the JWST/NIRCam observation and data are described in \citet{Shimakawa2024b}.
Fig.~\ref{fig2} also plots the 3D-HST sources at $z=$ 1.5--2.5 in the GOODS-S field \citep{Skelton2014} for reference, where their stellar masses and SFR$_\mathrm{SED}$ are obtained in the same manner by S24 (section~4.1), based on compilation data from \citet{Giavalisco2004,Liu2017,Luo2017,Nonino2009,Erben2005,Hildebrandt2006,Cardamone2010,Grogin2011,Koekemoer2011,Brammer2012,Wuyts2008,Retzlaff2010,Hsieh2012,Ashby2013}.

%%%%%%%%%%%%%%%%%%%%%%%%%%%%%%%%%%%%%%%%%%%%%%%%%%
\section{Resolving galaxy star formation and nuclear activities}
\label{s3}

\begin{figure}
\centering
\includegraphics[width=0.95\columnwidth]{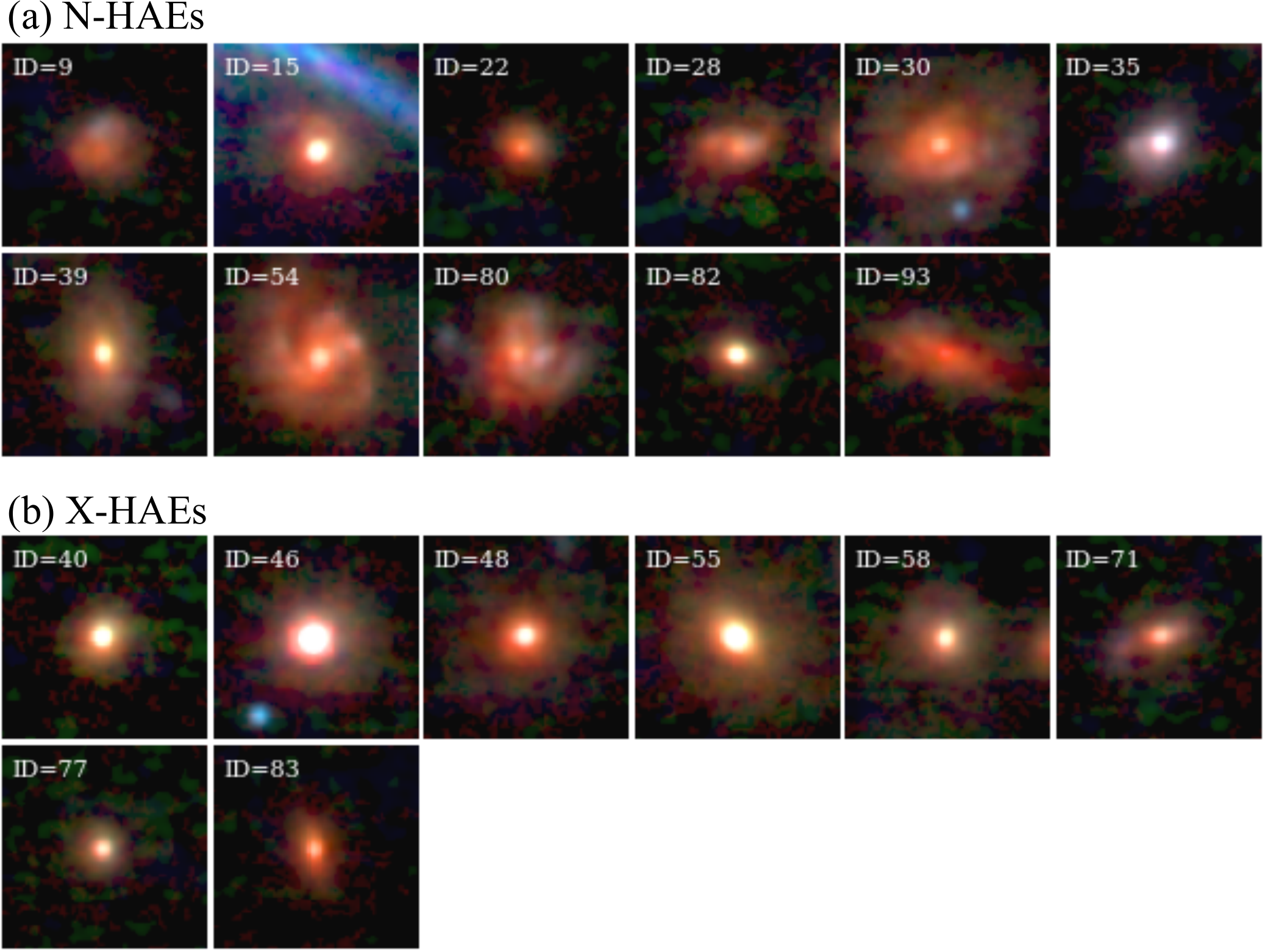}
\caption{
RGB colour postage stamps of (a) N-HAEs and (b) X-HAEs based on JWST/NIRCam F410M/F182M/F115W band images \citep{Shimakawa2024b,Perez-Martinez2024}.
}
\label{fig3}
\end{figure}

The remainder of this letter focuses on 19 massive HAEs with M$_\star>2\times10^{10}$ M$_\odot$ in the Spiderweb protocluster covered by the NIRCam imaging. 
Their RGB cutouts from F410M/F182M/F115W images are shown in Fig.~\ref{fig3}.
This section attempts to resolve star formation and nuclear activities in these N-HAEs and X-HAEs at a $\sim$ one kpc angular resolution using their composite Pa$\beta$ line images.

% figure 3

\begin{figure}
\centering
\includegraphics[width=0.9\columnwidth]{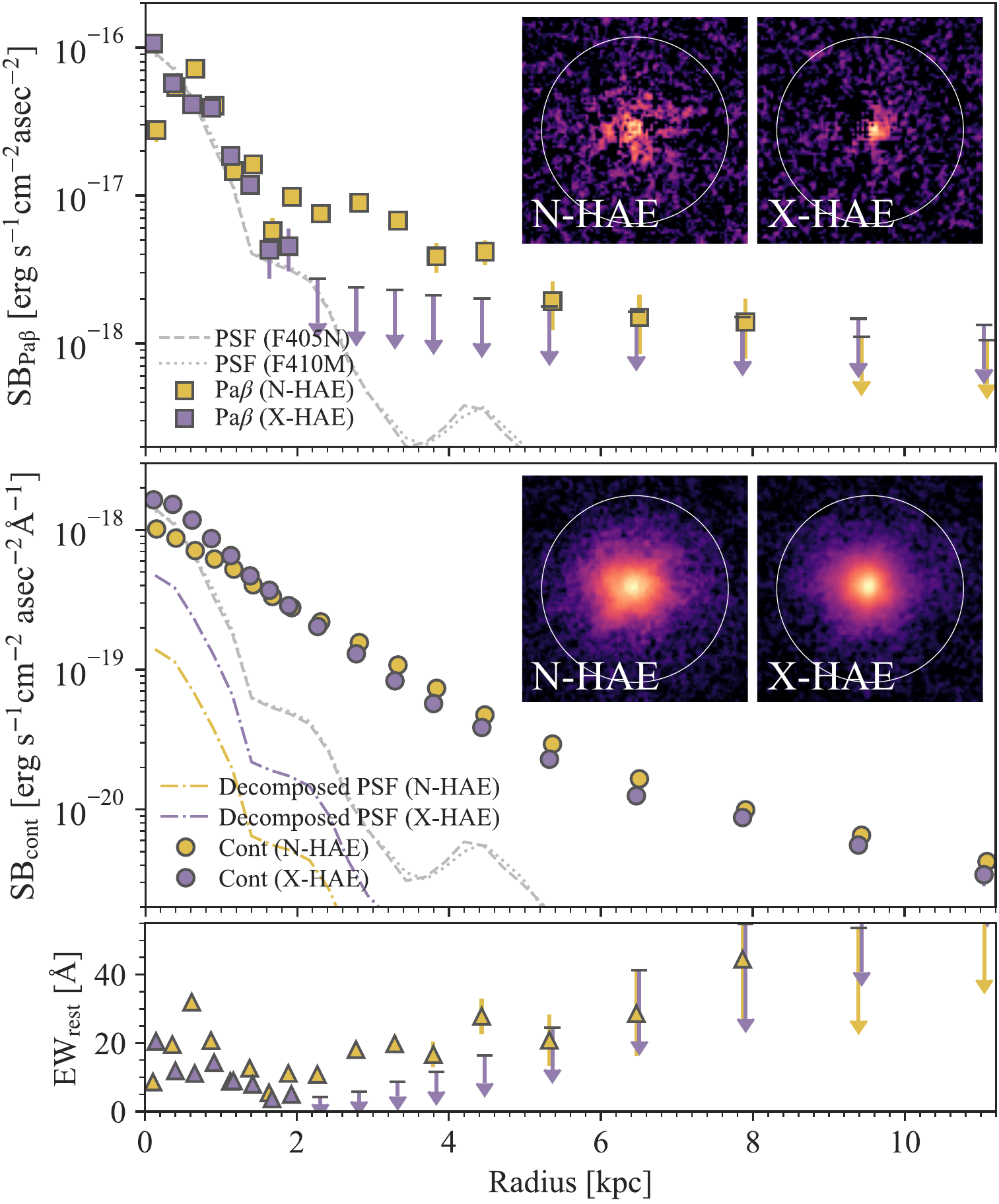}
\caption{
Radial profiles of Pa$\beta$ lines (top), rest-frame 1.3 $\mu$m continua (middle), and the rest-frame EWs (bottom). 
The radial profiles are based on the composite Pa$\beta$ and continuum images embedded in each panel, which are produced using the arcsinh stretch for the sake of visibility (the white circles indicate $r=$ 10 kpc). 
The yellow and magenta symbols indicate median radial profiles of N-HAEs and X-HAEs, respectively.
The colour regions depict $1\sigma$ uncertainties and inverse triangles indicate $2\sigma$ upper limits.
The grey dashed and dotted lines show composite PSFs in F405N and F410M, respectively.
The dash-dot lines are the possible PSF components decomposed from the continuum images (see Section~\ref{s4}).
}
\label{fig4}
\end{figure}

The resultant composite Pa$\beta$ and continuum images and their radial profiles are shown in Fig.~\ref{fig4}.
We here focus on the radial profiles owing to the lack of signal-to-noise ratios (SNRs) per pixel in the composite Pa$\beta$ line images.
They are based on the median stacking procedure described in \citet{Shimakawa2022} and $1\sigma$ errors of radial profiles are determined from standard deviations of the corresponding aperture or annulus on 500 random composite backgrounds in each band.
For the noise estimation, we randomly cutout images from background areas with similar weight values to individual HAEs based on the weight map. 
Matching weight values in the random background selection is critical because the imaging depth is not homogeneous across the survey area (Shimakawa et al. 2024b, section~2.2).
We do not rotate images in the stacking analysis to keep a consistent point spread functions (PSFs) within the samples.
The Pa$\beta$ surface brightness (SB$_\mathrm{Pa\beta}$) and stellar continuum at $\lambda_\mathrm{rest}=1.3$ $\mu$m (SB$_\mathrm{cont}$) within each annulus are estimated as follows:
\begin{eqnarray}
\mathrm{SB_{Pa\beta}} &=& \frac{1}{A}\frac{\Delta_\mathrm{NB}(f_\mathrm{NB}-f_\mathrm{BB})}{1-\Delta_\mathrm{NB}/\Delta_\mathrm{BB}}, \label{eq1}\\
\mathrm{SB_{cont}} &=& \frac{1}{A}\frac{f_\mathrm{BB}-f_\mathrm{NB}\cdot\Delta_\mathrm{NB}/\Delta_\mathrm{BB}}{1-\Delta_\mathrm{NB}/\Delta_\mathrm{BB}}, \label{eq2}
\end{eqnarray}
where A is an area of each annulus in arcsec$^2$, and $f_\mathrm{NB}$ and $f_\mathrm{BB}$ are flux densities within the corresponding annulus in the F405N and F410M bands, respectively.
We adopt filter bandwidths of $\Delta_\mathrm{NB}$=460~\AA\ and $\Delta_\mathrm{BB}$=4360~\AA. 
We also confirm that their PSFs are consistent with each other based on composite images of unresolved point sources (Fig.~\ref{fig4}).

% figure 4

Fig.~\ref{fig4} indicates that radial profiles of surface brightness are clearly different between N-HAEs and X-HAEs particularly at a radius of $r=$ 2--5 kpc.
The composite N-HAEs show disk star formation traced by Pa$\beta$ from $\sim2$ to $\sim8$ kpc, and a moderate increase in Pa$\beta$ EW is consistent with the inside-out growth observed in typical star-forming galaxies (e.g., \citealt{Nelson2012,Nelson2016,Shen2024}). 
On the other hand, we do not observe any clear sign of extended Pa$\beta$ component above $2\sigma$ levels in X-HAEs. 
Rather, there is an unresolved Pa$\beta$ emission in the centre, while their composite stellar radial profile is resolved (more extended) relative to the PSF.
The central point source dominates the entire Pa$\beta$ flux in X-HAEs, and it is reasonable to consider that it originates from X-ray AGNs rather than star formation \citep{Tozzi2022a}.
Thus, the comparison results of radial profiles in Pa$\beta$ and stellar continuum exhibit direct evidence of significantly lower star formation in X-HAEs than those in N-HAEs, which is claimed by S24 based on the multi-wavelength SED fitting.
However, it is currently unclear whether an offset Pa$\beta$ peak at $r=0.6$ kpc seen in N-HAEs is intrinsic or not: it could be due to stronger stellar absorption in the centre, the natures of the individuals, and measurement errors in object positions.
Because some N-HAEs may host Type 2 AGNs according to line diagnostics by \citet{Shimakawa2015,Perez-Martinez2023}, there may be partial point source contributions to their stacked Pa$\beta$ image. 

Then, total Pa$\beta$ fluxes are estimated to be $(11.4\pm0.2)\times10^{-18}$ erg~s$^{-1}$cm$^{-2}$ for N-HAEs and $(5.0\pm0.2)\times10^{-18}$ erg~s$^{-1}$cm$^{-2}$ for X-HAEs.
We calculate their SFR$_\mathrm{Pa\beta}$ through the \citet{Kennicutt1998} prescription with a conversion factor of H$\alpha$/Pa$\beta=17.5$ and assuming Pa$\beta$ extinctions of $A_\mathrm{Pa\beta}=$ 0.59 and 0.40 mag for N-HAEs and X-HAEs, respectively, based on the SED fitting for their median flux densities (S24).
A possible impact of Pa$\beta$ stellar absorption is ignored in this study (cf. EW=1--2~\AA\ according to \citealt{Seille2024}, corresponding to 5--14\% of the total flux in our sample). 
As a result, their dust-corrected SFR$_\mathrm{Pa\beta}$ are estimated to be $59\pm9$ and $<19$ M$_\odot$yr$^{-1}$ in N-HAEs and X-HAEs, respectively, which are consistent with those inferred from the SED fitting (Fig.~\ref{fig2}). 
We here adopt $2\sigma$ upper limit for X-HAEs as we do not detect a residual Pa$\beta$ component other than the point source in their composite Pa$\beta$ radial profile (Fig.~\ref{fig4}).
The total Pa$\beta$ flux is tentatively adopted for N-HAEs, although they could include AGN contributions. 
Multi-line information is required for individual sources to resolve their possible AGN impacts and better constrain their SFRs individually.

%%%%%%%%%%%%%%%%%%%%%%%%%%%%%%%%%%%%%%%%%%%%%%%%%%
\section{Impact of AGN feedback on star formation}
\label{s4}

\begin{figure}
\centering
\includegraphics[width=0.8\columnwidth]{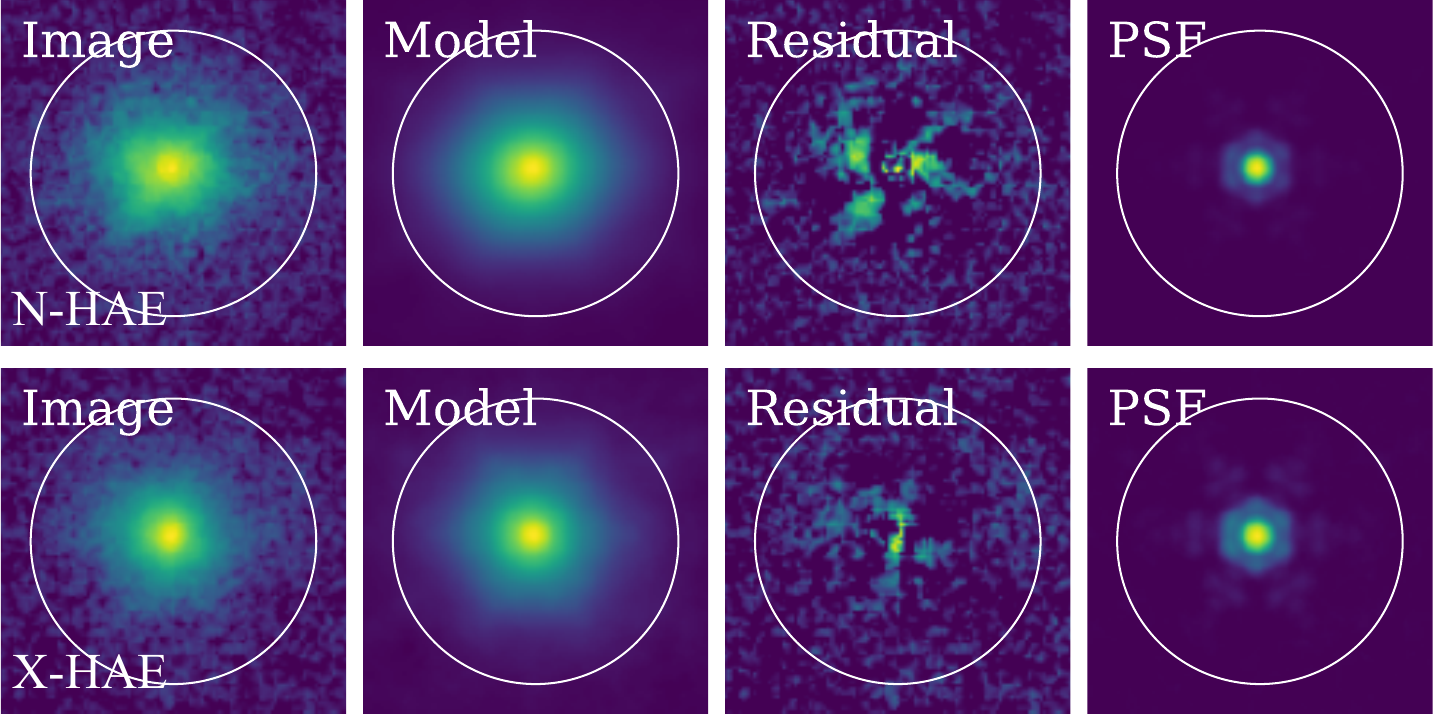}
\caption{
Model fitting to the continuum images (shown using the arcsinh stretch).
From above, the fitting results on the composite images of N-HAEs and X-HAEs are shown.
From left to right, each column represents the composite images, best-fit models, residual images, and decomposed PSFs, respectively, with a linear stretch in the same flux range (see also Fig.~\ref{fig4}).
}
\label{fig5}
\end{figure}

We find that star formation is significantly suppressed in massive HAEs hosting X-ray AGNs by comparing their spatially resolved Pa$\beta$ radial profile with those of the control HAE sample without X-ray counterparts.
Thus, this section discusses the obtained results in light of AGN feedback. 
Today, the black hole (BH) mass is considered to be a proxy for the total amount of AGN feedback energy, which is roughly proportional to the integral of BH accretion rates. 
The cumulative feedback energy budget is thought to be critical in regulating star formation quenching for massive galaxies in cosmological hydrodynamic simulations (e.g., \citealt{Terrazas2020,Wellons2023}). 
It should be borne in mind that this scenario is still debated \citep{Wang2023,Garza2024}.
However, as we cannot derive BH masses for our sample from the current data, we discuss AGN feedback instead by adopting their stellar potential values, $\phi_\star\equiv\mathrm{M}_\star/r_\mathrm{eff}$, as suggested by \citet{Terrazas2020,Piotrowska2022,Bluck2023,Bluck2024}. 
This is based on the assumptions that the BH mass is tightly correlated with the central velocity dispersion, that is, the gravitational potential (e.g., \citealt{Ferrarese2000,Gebhardt2000,Gultekin2009,Kormendy2013}), and the stellar potential approximates the latter quantity \citep[section~4.2]{Bluck2023}.
We estimate circularised effective radii ($r_\mathrm{eff}$) for our samples with {\sc galfit} (version 3.0.5, \citealt{Peng2010}) based on the F410M imaging data with the stacked PSF image ($\lambda_\mathrm{rest}=1.3$ $\mu$m).
In addition to a Sersic profile \citep{Sersic1963,Sersic1968}, a PSF model is included in the parameter fitting to consider a possible AGN contribution (Fig.~\ref{fig5}), and confirm larger effects of additional point source on the size measurements in X-HAEs compared to N-HAEs. 
For reference, point source contributions are estimated to be $f_\mathrm{PSF}=4\pm1\%$ in the stacked N-HAEs and $12\pm2\%$ in the stacked X-HAEs (see also Figs.~\ref{fig4}-\ref{fig5}). 

% figure 5

\begin{figure}
\centering
\includegraphics[width=0.9\columnwidth]{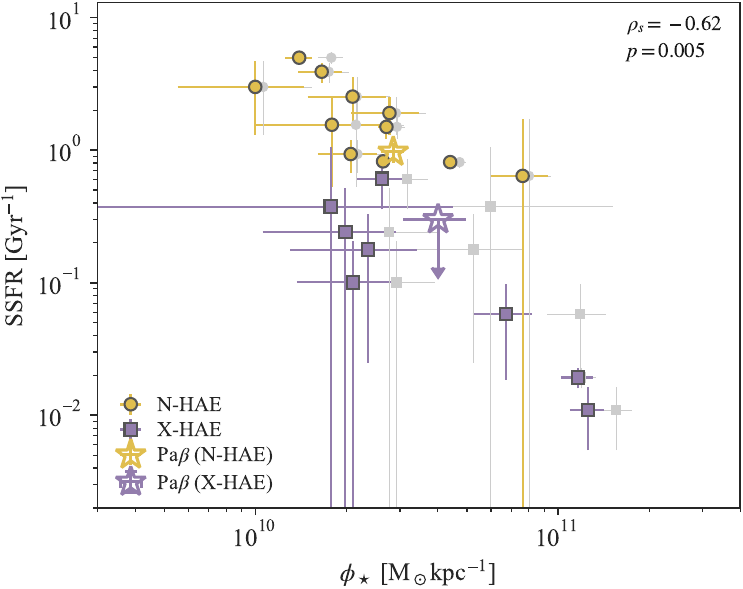}
\caption{
SSFRs versus stellar potentials ($\phi_\star$) for the selected HAEs with M$_\star>2\times10^{10}$ M$_\odot$.
The symbols are the same as in Fig.~\ref{fig2} but grey colours are based on size measurements using only Sersic components without additional point source contributions.
SSFRs of the composite N-HAEs and X-HAEs are based on dust-corrected SFR$_\mathrm{Pa\beta}$.
Spearman's rank correlation coefficient ($\rho_s=-0.62$) and $p$-values ($=0.005$) are appended in the top right corner.
}
\label{fig6}
\end{figure}

Fig.~\ref{fig6} shows the specific SFR (SSFR) against the stellar potential for our targets, suggesting that X-HAEs with low SSFRs tend to have higher stellar potentials compared to N-HAEs. 
The anti-correlation is also supported by Spearman's rank correlation test ($\rho_s=-0.62$ with $p$-values $=0.005$).
For the composite samples, we measure SSFRs from dust-corrected SFR$_\mathrm{Pa\beta}$ obtained in Section~\ref{s3}. 
Advanced mass concentrations in X-HAEs suggest that they may host more massive BHs and thus be injected by larger amounts of AGN feedback energies in their formation histories. 
Altogether, we conclude that X-HAEs may be at the stage of AGN quenching in the galaxy--BH co-evolution while emitting weak signs of AGNs in H$\alpha$+[N{\sc ii}], Pa$\beta$, and X-ray before permanent quenching.
At the moment it is restricted to Pa$\beta$ and continuum profiles, but resolving ionisation mechanisms, e.g., by observing multi lines with an integral field unit, may lead to greater understanding of these AGN host galaxies.

% figure 6

%%%%%%%%%%%%%%%%%%%%%%%%%%%%%%%%%%%%%%%%%%%%%%%%%%
\section{Conclusions}
\label{s5}

We study radial profiles of Pa$\beta$ and rest-frame NIR stellar continua for massive H$\alpha$-emitting galaxies that host X-ray AGNs (X-HAEs) and not (N-HAEs) in the Spiderweb protocluster at $z=2.16$, based on the NIRCam images and the source catalogue from our previous work (S24). 
The highlights of this letter are summarised as follows:
\begin{description}
    \item[--] We find that Pa$\beta$ radial profiles are clearly different between N-HAEs and X-HAEs in the sense that AGNs would dominate Pa$\beta$ line emissions in X-HAEs. 
    This provides direct evidence of low star formation in X-HAEs claimed by our previous study via multi-wavelength SED fitting (S24).
    \item[--] We also investigate the structural parameter, $\phi_\star$, which can be used as an observable proxy of the gravitational potential. 
    We then confirm that X-HAEs tend to have higher stellar potentials with lower SSFRs, further supporting the AGN quenching scenario to explain their low star formation in the outskirts of massive galaxies.
\end{description}
The Spiderweb protocluster is an ideal test bed for understanding the formation of the cluster red sequence at the peak epoch of cosmic star formation and BH growth. 
High-quality NIRCam images enable various detailed analyses that would address transitions from star-forming to red sequence galaxies and their environmental dependence by combining valuable archive data. 
Although we leave many of them to future research, this letter successfully illustrates some of such remarkable capabilities as an early report.

%%%%%%%%%%%%%%%%%%%%%%%%%%%%%%%%%%%%%%%%%%%%%%%%%%
\section*{Acknowledgements}

%We thank the anonymous referee for useful comments. 
This work is based on observations made with the NASA/ESA/CSA James Webb Space Telescope. 
The data were obtained from the Mikulski Archive for Space Telescopes at the Space Telescope Science Institute, which is operated by the Association of Universities for Research in Astronomy, Inc., under NASA contract NAS 5-03127 for JWST. 
The observation is associated with program \#1572 in cycle~1.
This work is in part based on observations taken by the 3D-HST Treasury Program (GO 12177 and 12328) with the NASA/ESA HST, which is operated by the Association of Universities for Research in Astronomy, Inc., under NASA contract NAS5-26555. 

This work is supported by a Waseda University Grant for Special Research Projects (2023C-590 and 2024R-057) and MEXT/JSPS KAKENHI Grant Numbers (23H01219 and 18H03717).
We would like to thank Editage (www.editage.com) for English language editing. 
TK acknowledges financial support from JSPS KAKENHI Grant Number 24H00002 (Specially Promoted Research by T. Kodama et al.).
HD, JMPM and YZ acknowledge financial support from the Agencia Estatal de Investigaci\'on del Ministerio de Ciencia e Innovaci\'on (AEI-MCINN) under grant (La evoluci\'on de los c\'umulos de galaxias desde el amanecer hasta el mediod\'ia c\'osmico) with reference (PID2019-105776GB-I00/DOI:10.13039/501100011033) and Agencia Estatal de Investigaci\'on del Ministerio de Ciencia, Innovaci\'on y Universidades (MCIU/AEI) under grant (Construcci\'on de c\'umulos de galaxias en formaci\'on a trav\'es de la formaci\'on estelar oscurecida por el polvo) and the European Regional Development Fund (ERDF) with reference (PID2022-143243NB-I00/10.13039/501100011033).
JMPM acknowledges funding from the European Union's Horizon-Europe research and innovation programme under the Marie Sk{\l}odowska-Curie grant agreement No 101106626.
C.D.E. acknowledges funding from the MCIN/AEI (Spain) and the NextGenerationEU/PRTR (European Union) through the Juan de la Cierva-Formacion program (FJC2021-047307-I).
YZ acknowledges the support from the China Scholarship Council (202206340048), and the National Science Foundation of Jiangsu Province (BK20231106).
This work made extensive use of the following tools, {\sc numpy} \citep{Harris2020}, {\sc matplotlib} \citep{Hunter2007}, {\sc topcat} \citep{Taylor2005}, {\sc astopy} \citep{AstropyCollaboration2013, AstropyCollaboration2018,AstropyCollaboration2022}, and {\sc pandas} \citep{Reback2022}.

%%%%%%%%%%%%%%%%%%%%%%%%%%%%%%%%%%%%%%%%%%%%%%%%%%
\section*{Data Availability}

The targets are selected from the publicly-available source catalogues in \citet{Shimakawa2018b,Shimakawa2024}. 
The original science frames are stored in and can be retrieved from the archive system termed the Barbara A. Mikulski Archive for Space Telescopes (MAST)\footnote{\doi{10.17909/vx25-q902}}.



\bibliographystyle{mnras}
\bibliography{spwb}

%\begin{thebibliography}{99}
%\end{thebibliography}

%%%%%%%%%%%%%%%%%%%%%%%%%%%%%%%%%%%%%%%%%%%%%%%%%%

% Don't change these lines
\bsp	% typesetting comment
\label{lastpage}
\end{document}